\documentclass[bibyear]{aa}

\usepackage{graphicx}
\usepackage{amsmath}
\usepackage{epsfig}
\usepackage{txfonts}

\begin{document}

\def\teff{$T_{\rm eff}$}
\def\logg{\log g}
\def\kms{km\,s$^{-1}$}
\def\msun{$M_{\sun}$}
\def\rsun{$R_{\sun}$}
\def\d{d$^{-1}$}

\newcommand\he[2]{He\,\textsc{#1}\,$\lambda$\,{#2}}
\newcommand\n[2]{N\,\textsc{#1}\,$\lambda$\,{#2}}

   \title{Short-term spectroscopic variability of Plaskett's star\thanks{Based on observations collected at the Observatoire de Haute Provence (OHP, France).}}

   \author{M. Palate
          \and
          G. Rauw
          }

   \institute{Institut d'Astrophysique et de G\'eophysique, 
   Universit\'e de Li\`ege, B\^at. B5c, All\'ee du 6 Ao\^ut 17, 4000 Li\`ege, Belgium\\}
	

   \date{Received 10 February 2014 / Accepted 30 September 2014}

	\abstract
   {Plaskett’s star (HD\,47129) is a very massive O-star binary in a post Roche-lobe overflow stage. {\it CoRoT} observations of this system revealed photometric variability with a number of frequencies.}
   {The aim of this paper is to characterize the variations in spectroscopy and investigate their origin.}
   {To sample its short-term variability, HD\,47129 was intensively monitored during two spectroscopic campaigns of six nights each. The spectra were disentangled and Fourier analyses were performed to determine possible periodicities and to investigate the wavelength dependence of the phase constant and the amplitude of the periodicities.}
   {Complex line profile variations are observed. Frequencies near 1.65, 0.82, and 0.37\,\d\ are detected consistently in the \he{i}{4471}, \he{ii}{4542}, and \n{iii}{4510-4518} lines. These frequencies are consistent with those of the strongest signals detected in photometry. The possibilities that these variations stem from pulsations, a recently detected magnetic field or tidal interactions are discussed.}
   {Whilst all three scenarios have their strengths, none of them can currently account for all the observed properties of the line profile variations.}
	\authorrunning{Palate and Rauw}
	\titlerunning{Plaskett's star}

   \maketitle

\section{Introduction}
The massive binary nature of HD\,47129 was first reported by Plaskett (\cite{Plaskett}), who inferred a total mass of 138.9 \msun, the highest ever observed at that time. He described the binary as consisting of two early-type stars: an Oe5 primary and a fainter secondary with weak and broad lines. For almost a century, HD\,47129, also known as Plaskett's star, has been the target of many studies, with tremendous progress in the past twenty years (e.g.\ Bagnuolo et al. \cite{Bagnuolo}, Wiggs \& Gies \cite{Wiggs}, Linder et al.\ \cite{Linder06}, \cite{Linder08}, Mahy et al.\ \cite{Mahy}, Grunhut et al.\ \cite{Grunhut}). The masses of the components have been revised downwards to $m_1\,\sin^3{i} = 45.4$ for the primary and $m_2\,\sin^3{i} = 47.3$ \msun\ for the secondary, leading to a mass ratio of about $1.05\pm0.05$ (Linder et al.\ \cite{Linder08}). Unfortunately, Plaskett's star is a non-eclipsing binary, preventing a precise determination of its orbital inclination. The latter was estimated from polarimetry to $71 \pm 9$\degr\ (Rudy \& Herman \cite{Rudy}). Plaskett's star is usually considered a member of the Mon\,OB2 association. However, Linder et al.\ (\cite{Linder08}) find a discrepancy between the luminosity and the dynamical masses of the two components of the system, which could be solved by assuming a greater distance for the star. 

In many respects, HD\,47129 offers a textbook example of evolutionary effects and the effects of interactions in massive binaries. Linder et al. (\cite{Linder08}) used a disentangling technique, based on the method of Gonz\'alez \& Levato (\cite{Gonzalez}), on high-resolution optical spectra to derive the spectral types O8 III/I and O7.5 V/III for the primary and secondary stars, respectively. The broad and shallow absorption lines of the secondary star suggest that this star rotates rapidly ($v\sin{i} \sim 300$\kms), whereas the much sharper absorption lines of the primary indicate a projected rotational velocity of $v\sin{i} \sim75$\kms\ (Linder et al. \cite{Linder08}). Using the CMFGEN model atmosphere code (Hillier \& Miller \cite{HM}) to analyse the disentangled spectra, Linder et al.\ (\cite{Linder08}) show that the atmosphere of the primary has an enhanced N and He abundance and a depletion of C. The secondary atmosphere is possibly depleted in N. These results indicate that the binary system is in a post-Roche lobe-overflow evolutionary stage where matter and angular momentum have been transferred from the primary to the secondary. 

Because of its high rotation speed, the wind of the secondary star could be rotationally flattened. Such a situation would affect the properties of the wind interaction zone in this binary. This was confirmed by the studies of the H$\alpha$ emission region by Wiggs \& Gies (\cite{Wiggs}) and Linder et al. (\cite{Linder08}), as well as by the study of the X-ray emission of the system by Linder et al. (\cite{Linder06}). An alternative explanation for the origin of the equatorial wind of the secondary could be magnetic confinement. Indeed, Grunhut et al. (\cite{Grunhut}) have recently reported the presence of a magnetic field in Plaskett's star. In the framework of the magnetism in massive stars (MiMeS) survey, they detected Zeeman signatures from the rapidly rotating secondary in high-resolution Stokes V spectra. The strength of the highly organized field has to be at least $2850 \pm 500$ G, and variations compatible with rotational modulation of an oblique field were also found (Grunhut et al.\ \cite{Grunhut}).

	\begin{table*}
		\caption{Summary of the time sampling of the observing campaigns.}
		\label{tab:obs}
		\centering
			\begin{tabular}{l c c c c c c c c}
			 \hline\hline
			  Campaign & Starting and ending dates & Starting and ending& n 	& <S/N> & $\Delta T$ & $<\Delta t>$ & $\Delta \nu_{nat}$ & $\nu_{max}$ \\
									& (HJD-2450000)							&  phases & 		& 				&   (days)	 &   (hours)		& (days)$^{-1}$ & (days)$^{-1}$ \\
				\hline
				2009 & 5174.415 - 5179.665 						& 0.47 - 0.84 & 68 	& 180 		& 5.25 				& $0.40$ & $1.90~10^{-1}$ & 119 \\
				2010 & 5539.454 - 5543.657 						& 0.83 - 0.12 &	50	&	146 		& 4.20 				&	$0.55$ & $2.38~10^{-1}$ & 87.7 \\
				\hline
			\end{tabular}
			\tablefoot{Note that for both campaigns, there are no observations for one night due to bad weather. n is the total number of spectra, <S/N> indicates the mean signal-to-noise ratio, $\Delta T$ indicates the total time between the first and last observations, $\Delta t$ provides the average time interval between two consecutive exposures of a same night. The last two columns indicate
the natural width of the peaks in the periodogram $\Delta \nu_{nat}= \Delta T^{-1}$ and $\nu_{max} = (2<\Delta t>)^{-1}$ provides an  indication of the highest frequency that can possibly be sampled with our time series.}
	\end{table*}	

Mahy et al. (\cite{Mahy}) studied the CoRoT (Convection, Rotation, and planetary Transits, Baglin et al. \cite{Baglin}; Auvergne et al. \cite{Auvergne}) light curve of Plaskett's star and extracted 43 significant frequencies. Among these frequencies there are three major groups: 0.823 \d\ and six harmonics, the orbital frequency 0.069 \d\ and two harmonics, as well as two frequencies at 0.368 \d\ and 0.399 \d. The latter two could be related to the rotation period of the secondary. 

The discovery of the 0.823\,\d\ frequency by Mahy et al.\ (\cite{Mahy}) prompted us to perform a spectroscopic monitoring of the system, to constrain the origin of this variation. In this paper, we report the results of this study. The paper is organized as follows. Sect.\,\ref{sect:obs} describes the observations, data reduction and disentangling treatment. In Sect.\,\ref{sect:var}, the line profile variations are discussed and the results of the Fourier analyses are given. Finally, in Sect.\,\ref{sect:scen}, we discuss the strengths and weaknesses of three possible scenarios and summarize our results and conclusions.

\section{Observations and data reduction \label{sect:obs}}

	\subsection{OHP spectroscopy}

Spectroscopic time series of HD\,47129 were obtained during two six-nights observing campaigns in December 2009 and December 2010 at the Observatoire de Haute Provence (OHP, France). We used the Aur\'elie spectrograph fed by the 1.52\,m telescope. The spectrograph was equipped with a 1200 lines\,mm$^{-1}$ grating  blazed at 5000\,\AA\ and a CCD EEV42-20 detector with $2048 \times 1024$ pixels of 13.5\,$\mu$m$^2$. Our set-up covered the wavelength domain from 4460 to 4670\,\AA\ with a resolving power of 20000. Typical exposure times were 20 -- 30 minutes, depending on the weather conditions, and a total of 118 spectra were collected. To achieve the most accurate wavelength calibration, Th-Ar lamp exposures were taken regularly over each observing night (typically once every 90 minutes).

The data were reduced using the MIDAS software provided by ESO. The normalization was done self-consistently using a series of continuum windows. Table \ref{tab:obs} provides a summary of our observing campaigns and the characteristics of the sampling.

	\subsection{Disentangling}\label{disentangling}
	
\begin{figure}[ht!]
		\resizebox{\hsize}{!}{\includegraphics{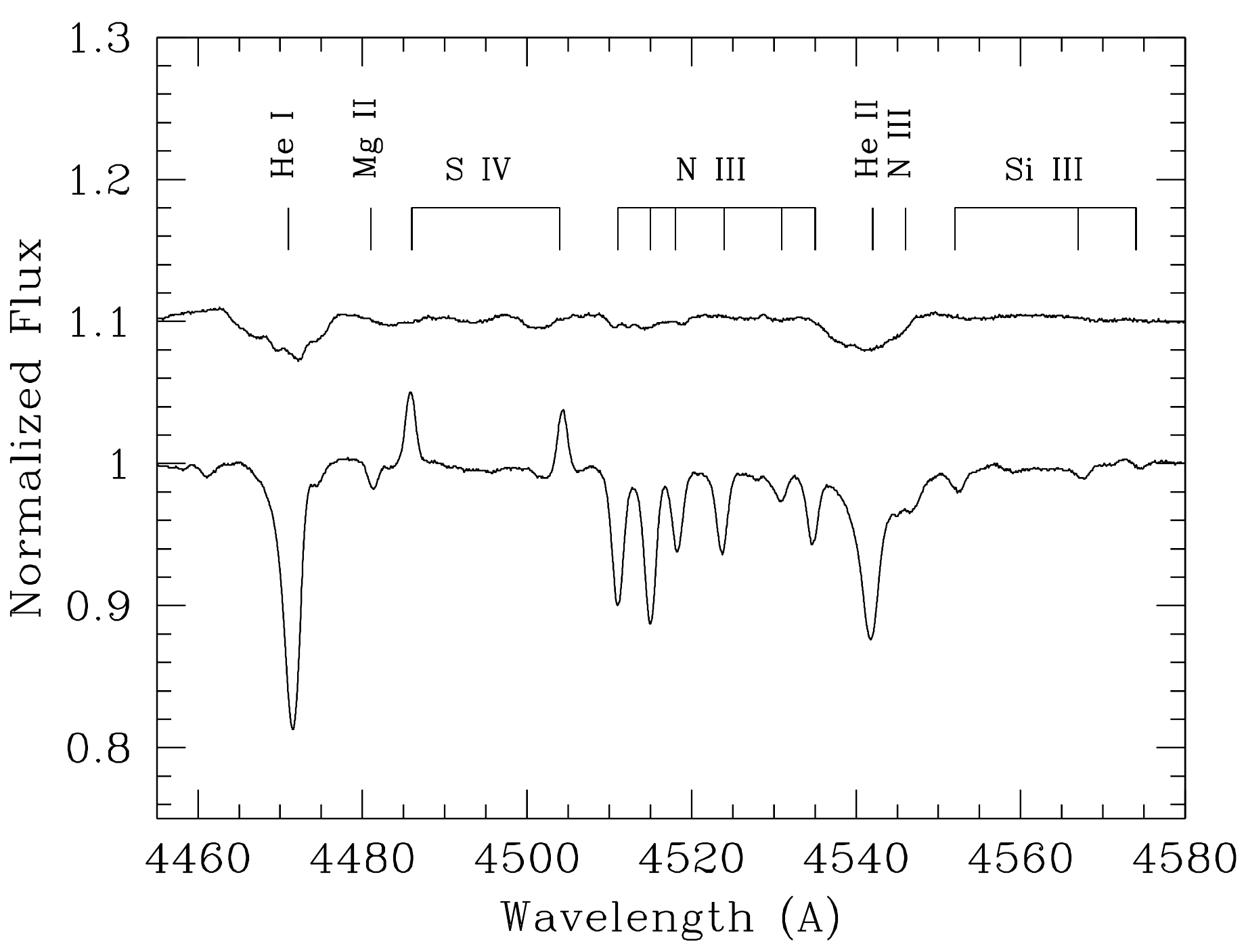}}
		\caption{Mean disentangled spectra of the primary and secondary (shifted upwards by 0.1 continuum units). Important spectral lines are identified by the labels.}
		\label{fig:disent}
\end{figure}

The goal of the present paper is to study and interpret the short-term line profile variability of Plaskett's star. Because of the Doppler shifts and line blending induced by the orbital motion in such a binary system, it is not trivial to know which star triggers the variations. The first step in our analysis was thus to disentangle the spectra of the primary and secondary components. For this purpose, we applied our code based on the Gonz\'alez \& Levato (\cite{Gonzalez}) technique to the data of the two observing campaigns simultaneously. This method allows to derive the radial velocities as well as the individual spectra of the components. However, because of the very broad and shallow lines of the secondary, its radial velocities (RVs) obtained with the disentangling were sometimes not well constrained. To avoid artefacts, we thus fixed the RVs of the secondary to the theoretical values derived from the orbital solution of Linder et al.\ (\cite{Linder08}). The situation for the primary was better, as we recovered RVs that were in good agreement with the orbital solution of Linder et al.\ (\cite{Linder08}): the dispersions of the RVs about the latter orbital solution were 11 and 6\,km\,s$^{-1}$ for the He\,{\sc i} $\lambda$\,4471 and He\,{\sc ii} $\lambda$\,4542 lines, respectively. There were very little differences between the disentangled spectrum of the primary obtained with either the primary RVs being fixed to the orbital solution or allowing them to vary. In the subsequent analysis of the line profile variability, we thus adopted the mean spectra as reconstructed keeping the RVs of the primary fixed to the orbital solution of Linder et al.\ (\cite{Linder08}). The reconstructed spectra of the individual components are displayed in Fig.\,\ref{fig:disent}. These individual spectra are important to correctly interpret the results of our line profile variability analysis. 
For instance, Fig.\,\ref{fig:disent} reveals that the only secondary line that has at least some parts of its profile that are not blended with primary lines at some orbital phases is \ion{He}{i} $\lambda$\,4471. The latter line displays weak emission humps that were already noted by Linder et al.\ (\cite{Linder08}).

\begin{figure*}[ht!]
\begin{center}
		\resizebox{!}{10cm}{\includegraphics{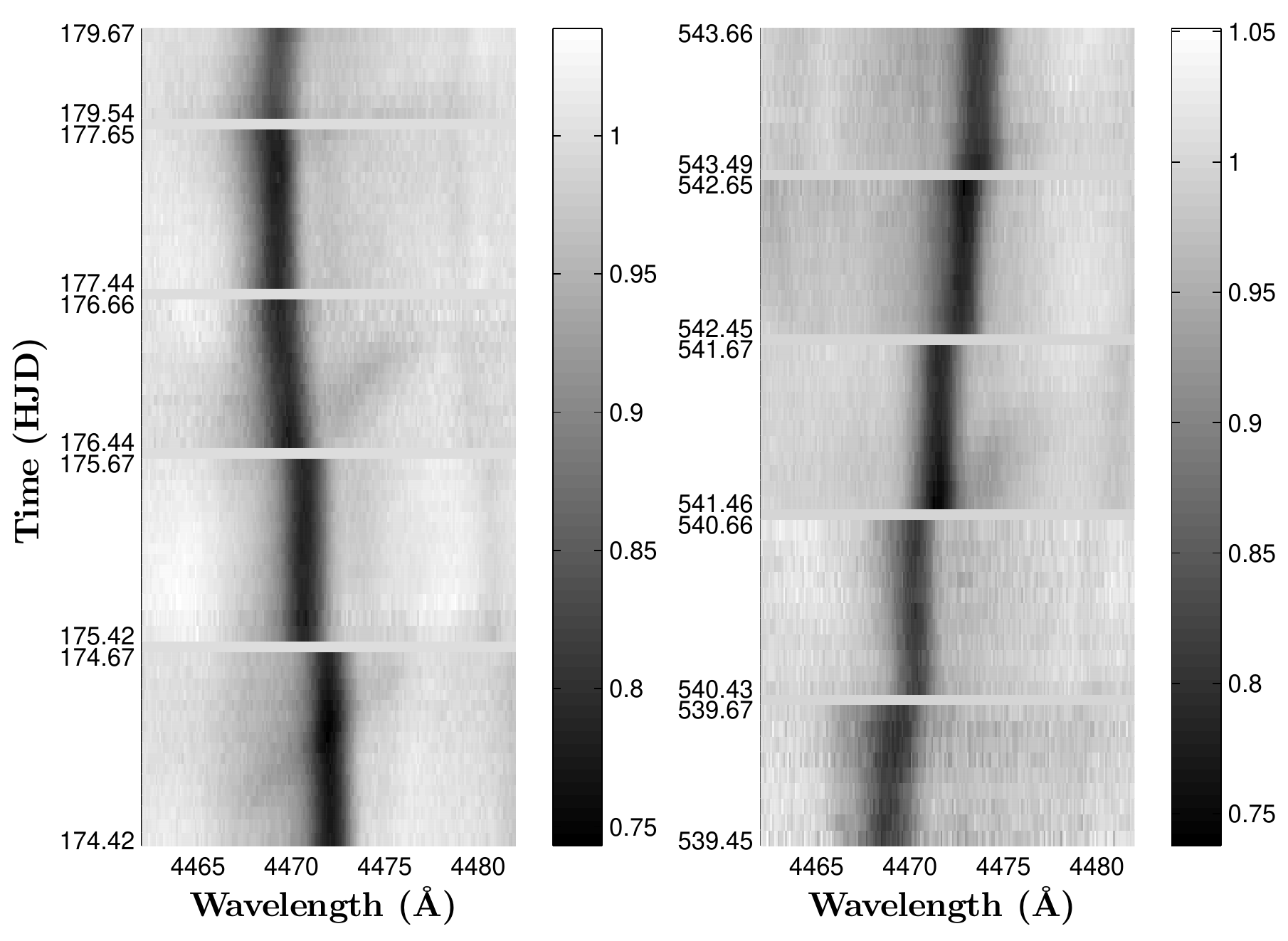}}
\end{center}

\begin{center}
		\resizebox{!}{10cm}{\includegraphics{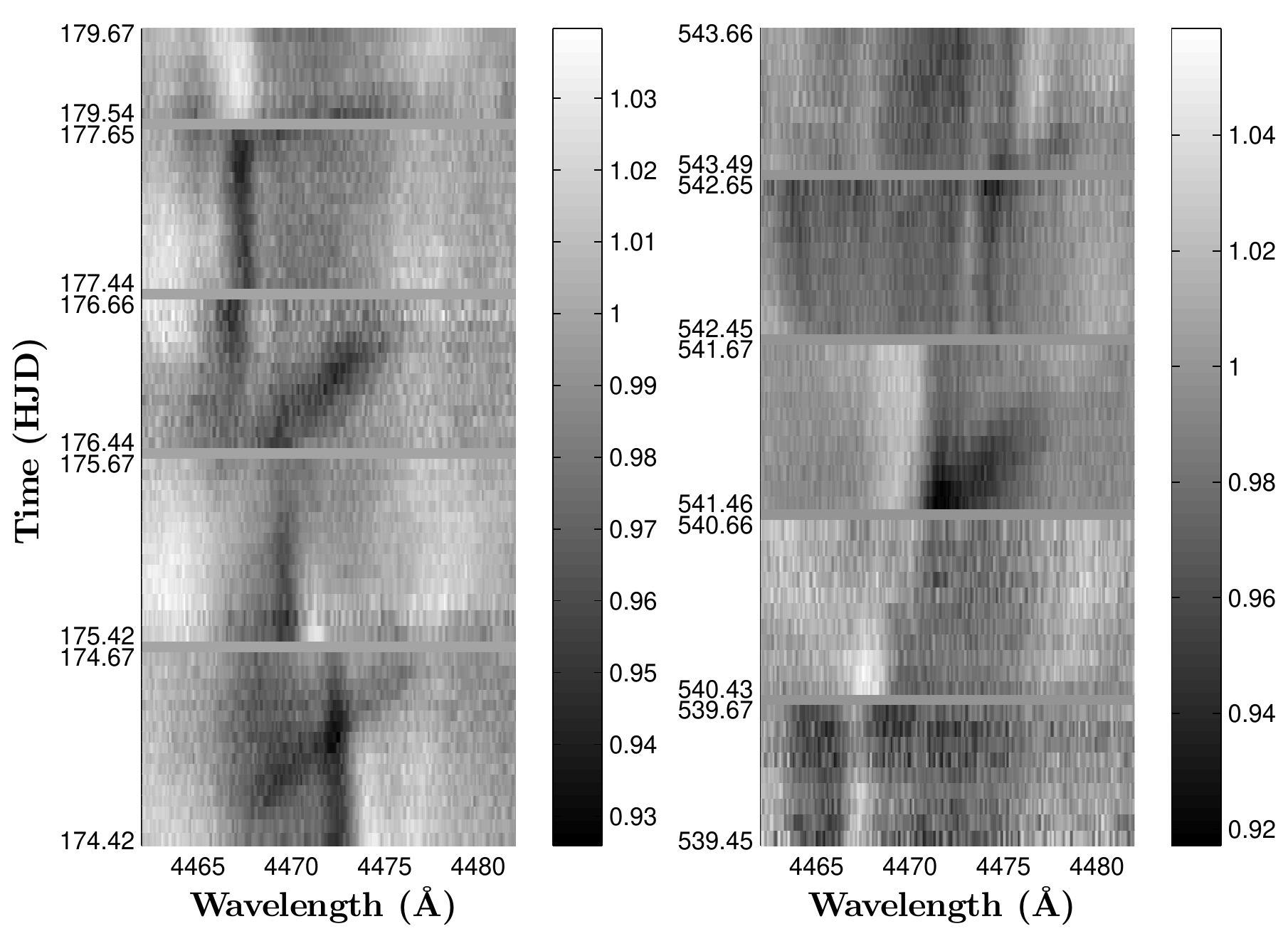}}
\end{center}

		\caption{Line-profile variations in the \ion{He}{i} $\lambda$\,4471 line in the 2009 (left) and 2010 (right) campaigns. The top panels show the dynamic spectra of the observed data in the heliocentric frame of reference. The bottom panels yield the secondary dynamic spectra. The labels on the vertical axis indicate the date of the first and last observation of each night in the format HJD-2455000.}
		\label{fig:dynsec}
\end{figure*}
\section{Variability analysis \label{sect:var}}
\subsection{\ion{He}{i} $\lambda$\,4471}
The top panels of Fig. \ref{fig:dynsec} display the dynamic spectra of the observed data in the heliocentric frame of reference (hereafter called the raw dynamic spectra), whilst the bottom panels illustrate the dynamic spectra after subtracting the mean disentangled primary spectrum and shifting the data into the secondary's frame of reference (hereafter called the secondary dynamic spectra).
 
The raw dynamic spectra are dominated by the orbital motion of the relatively sharp primary line. These figures show which parts of the core of the secondary absorption are most affected by blends with the primary. We note that the observed position of the primary line slightly deviates from the position expected from the orbital solution. Another feature that can be seen on the first and third night of the 2009 campaign as well as on the third night of the 2010 data is a discrete depression that moves from the blue to the red wing of the broad secondary absorption.

\begin{figure*}[ht!]
\begin{minipage}{6.6cm}
\includegraphics[width=6.6cm]{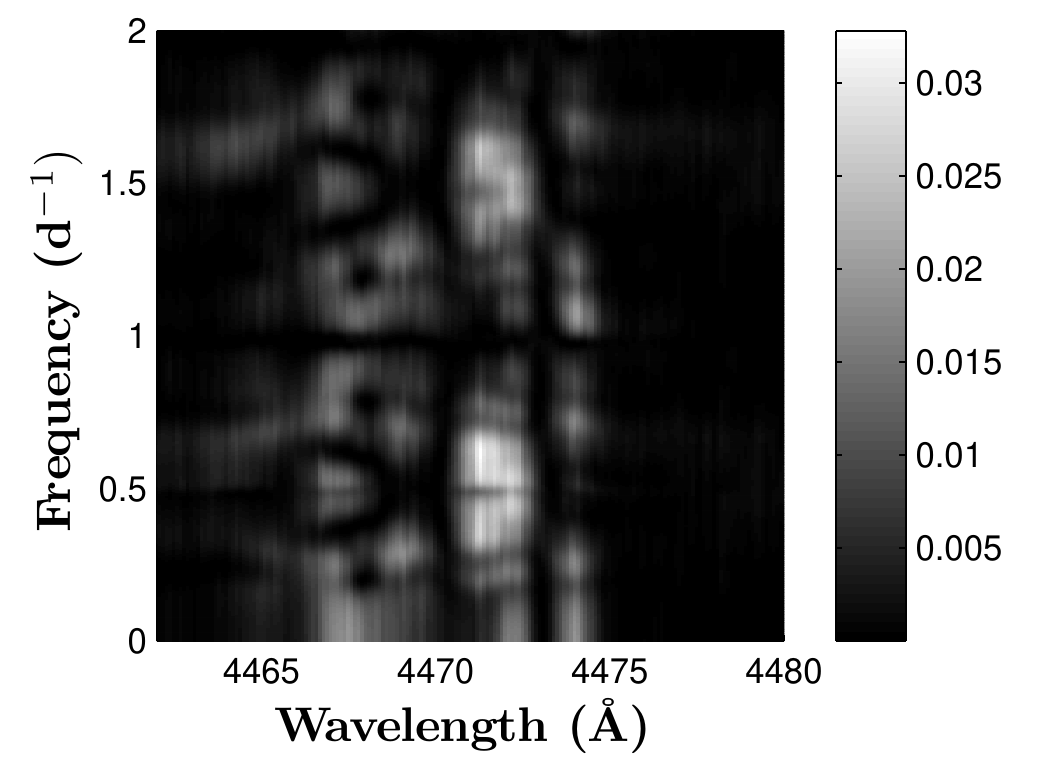}
\includegraphics[width=6.6cm]{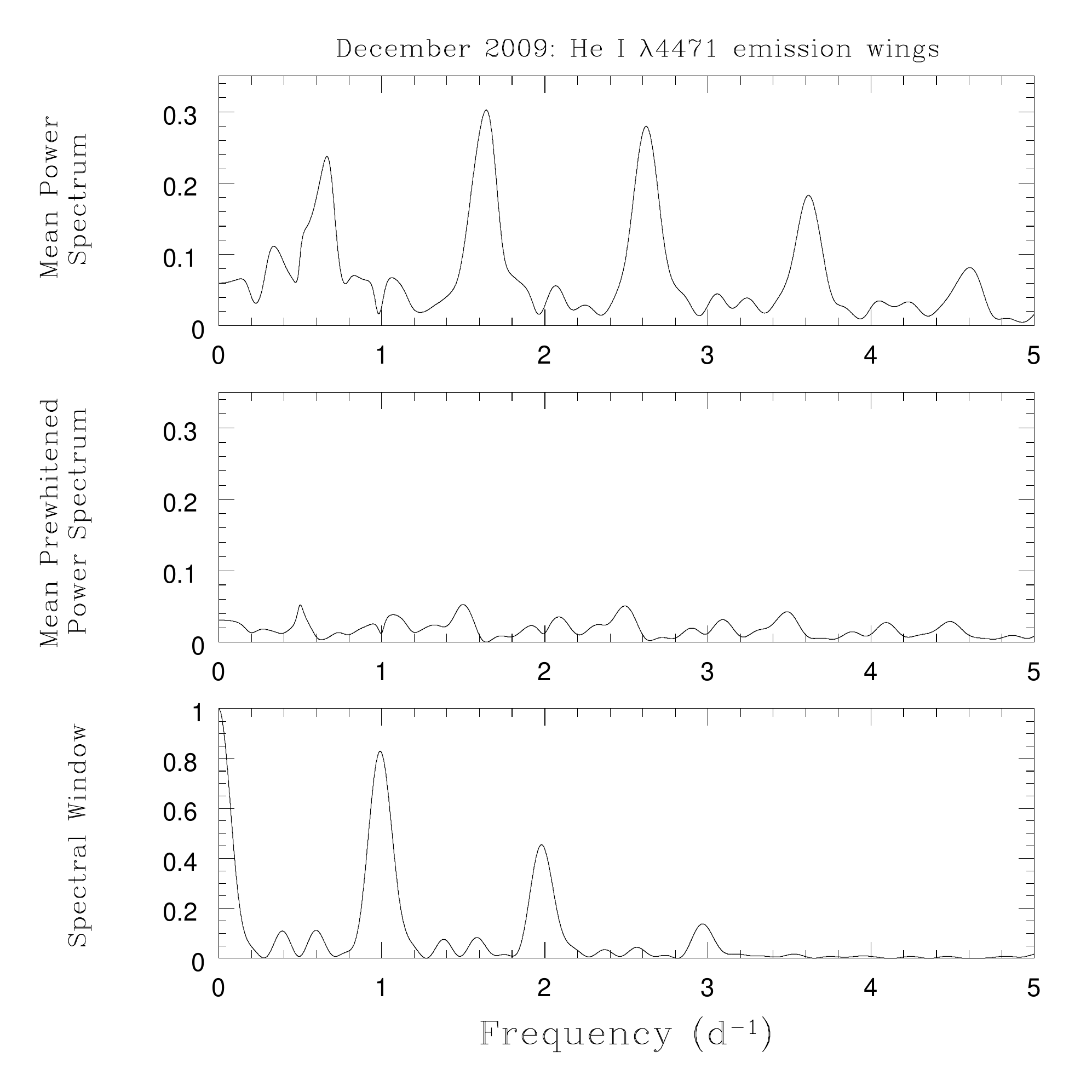}
\end{minipage}
\begin{minipage}{6.6cm}
\includegraphics[width=6.6cm]{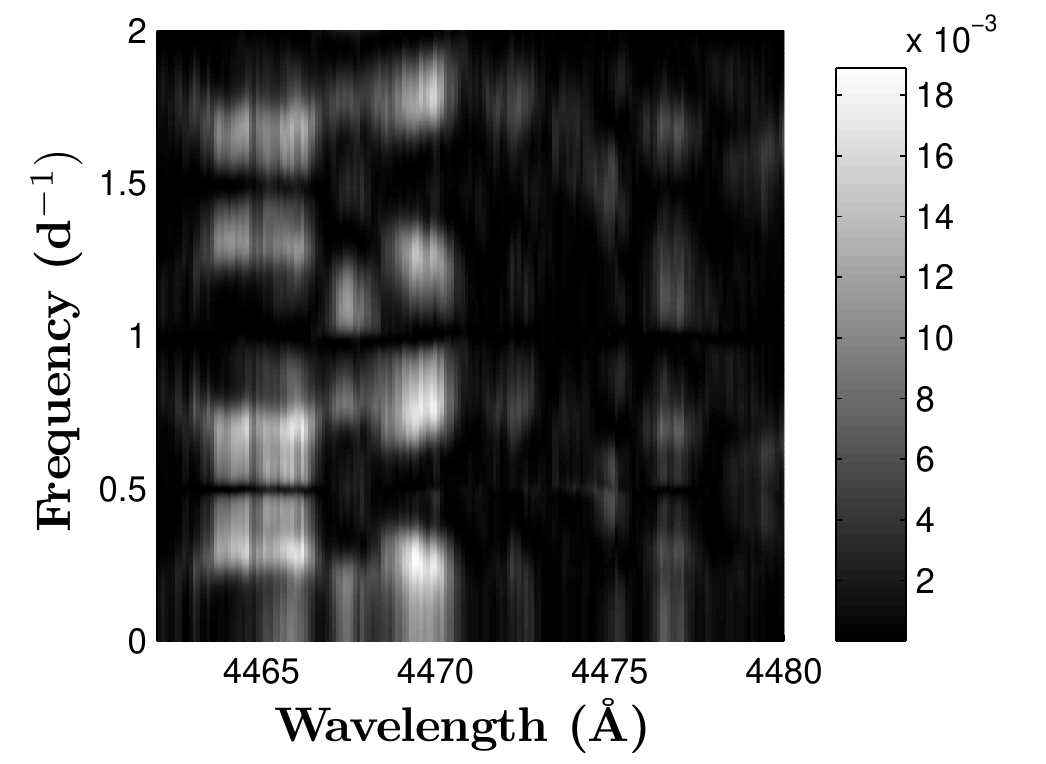}
\includegraphics[width=6.6cm]{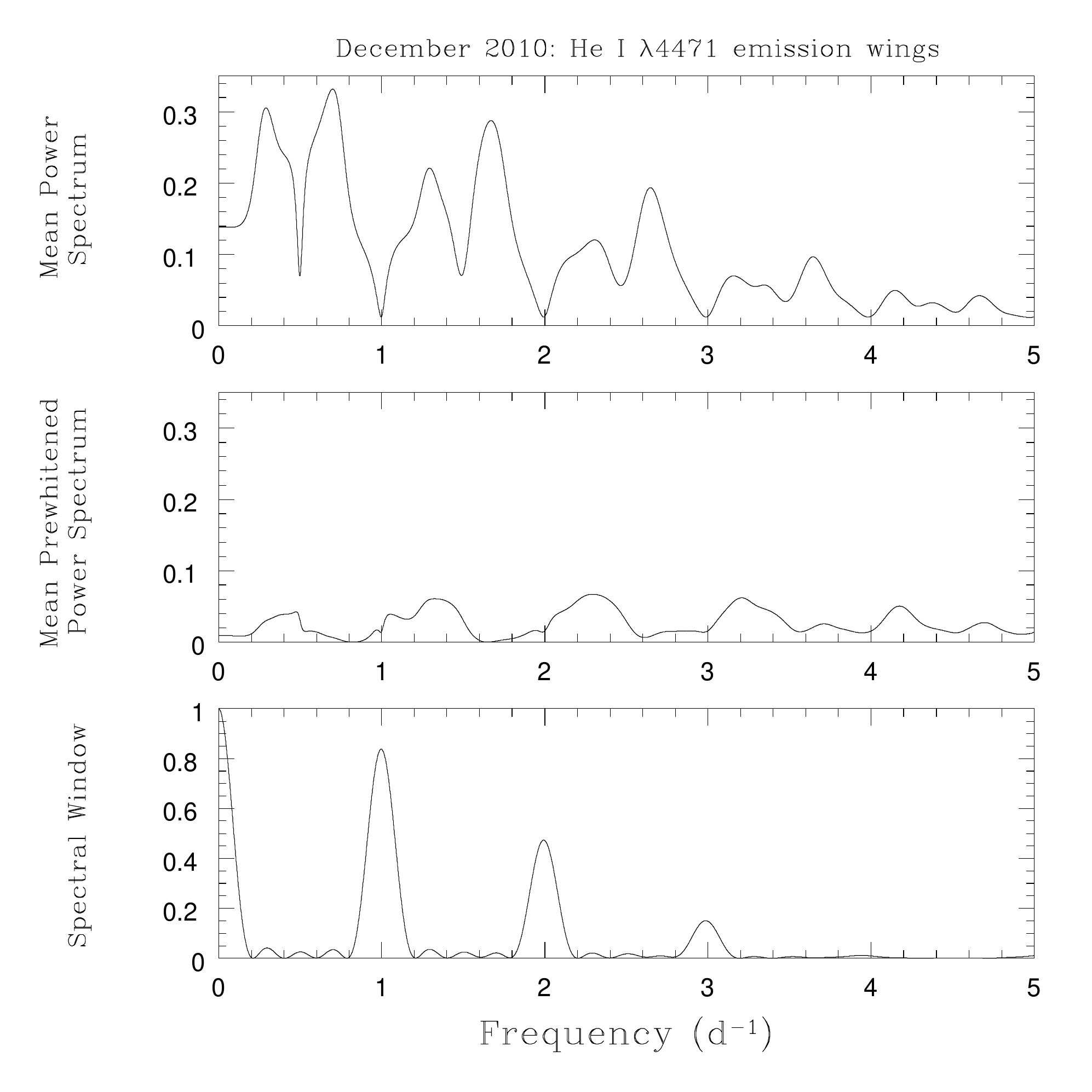}
\end{minipage}
\begin{minipage}{5cm}
		\caption{Fourier power spectra of the 2009 (left) and 2010 (right) time series of the secondary's \ion{He}{i} $\lambda$\,4471 line. The top panels of each row show the full 2D periodograms (between 0 and 2\,d$^{-1}$), whilst the second and third panels illustrate the mean periodogram over the emission humps, respectively before and after prewhitening (see text). The spectral window is shown in the lower panel.}
		\label{fig:TF4471}
\end{minipage}
\end{figure*}

The secondary dynamic spectra reveal another feature that is more difficult to distinguish on the raw dynamic spectra. The strengths of the blue (4462 -- 4465\,\AA) and red (4475 -- 4480\,\AA) emission humps that flank the secondary's absorption (see Fig.\,\ref{fig:disent}) undergo a strong modulation. The latter seems correlated with the presence of the discrete depression pointed out hereabove. Indeed, when the redwards moving discrete depression is observed, the emission humps are at their low level. The deviations between the observed and the theoretical positions and shapes of the primary profile lead to residuals when subtracting the mean primary spectrum that contaminate the secondary dynamic spectra between about 4466 and 4475\,\AA. 

The dynamic spectra reveal that the variations extend over a wider wavelength range than the widths of the primary lines. We thus conclude that the secondary very probably displays variations, although, at this stage, we cannot exclude that the primary is also variable, especially in view of the residuals after subtracting the mean primary profile.

To further characterize the variations in the \ion{He}{i} $\lambda$\,4471 line, we performed a 2D Fourier analysis (Rauw et al.\ \cite{Rauw}) based on the periodogram method of Heck et al.\ (\cite{Heck}) refined by Gosset et al. (\cite{Gosset}). This method is explicitly designed to account for the uneven sampling of the time series.

Applying our Fourier analysis to the time series of the observed spectra, without shifting them into the frame of reference of either star and/or subtracting the spectrum of the primary, results in power spectra dominated by low frequencies due to the orbital motion of the primary and, to a lesser extent, secondary lines. To get rid of these orbital frequencies and analyse the genuine short-term line profile variations, we have subtracted the mean disentangled spectrum of the primary shifted to the appropriate RV from the observed spectra. The resulting difference spectra were then further shifted into the secondary's frame of reference. Therefore, we obtained the ``individual spectra'' of the secondary in the frame of reference of this star for each observation. The resulting 2D periodograms of the corresponding 2009 and 2010 time series are shown in Fig.\,\ref{fig:TF4471}.

To start, we focus on the emission humps (4462 -- 4465 \& 4475 -- 4480\,\AA). We find that the mean periodogram of each observing campaign is dominated by a frequency near 1.65\,d$^{-1}$ and its aliases. The error on the centre of a peak in the periodogram can be estimated as $\Delta\nu = 0.1/\Delta T$\footnote{This corresponds to an uncertainty on the period of $\Delta P = 0.1\times\dfrac{P^2}{\Delta T}$. }. Here $\Delta T$ is the time interval between the first and the last observations of the time-series (see Table\,\ref{tab:obs}) and we assume that the uncertainty on the peak frequency amounts to 10\% of the peak width. The uncertainties are of 0.02\,\d\ for the campaigns separately and of $3\times10^{-4}$\,\d\ when both campaigns are considered simultaneously\footnote{We caution though that the sampling of the combined dataset is very odd, leading to a large number of closely spaced aliases, and the effective uncertainties on the peak position are thus significantly larger than estimated here.}. However, in the present case, these estimates are quite optimistic because they neither account for the signal to noise ratio of the observations nor for the impact of the treatment of the data prior to the Fourier analysis. Therefore, we estimate an overall error of $\pm 0.05$\,\d\ on the peak values.

Because of this rather large error on the position of the peak, some caution is needed when comparing the results of our present analysis with those of the {\it CoRoT} light-curve analysis of Mahy et al.\ (\cite{Mahy}). The latter authors reported a large number of frequencies and some of the peaks in our periodograms could be blends of two or more frequencies. Indeed, our spectroscopic time series are not sufficiently long to distinguish closely spaced frequencies. Moreover, our time series are heavily affected by 1-day aliasing, which was absent in the {\it CoRoT} data. For instance, the 1-day aliasing introduces some coupling between frequencies near 0.4\,\d\ and those near 1.6\,\d. Nevertheless, we note that our 1.65\,\d\ peak nicely fits the 1.646\,\d\ frequency of the strongest signal found in the {\it CoRoT} data. Table \ref{tab:corot} lists the frequencies from the {\it CoRoT} data that are potentially consistent with the frequencies found in the current analysis. 

To check whether the photometric frequencies can account for the periodograms of our data, we used a prewhitening technique (Rauw et al.\ \cite{Rauw}). In this way, we found that the 2009 periodogram can be prewhitened with a single frequency (1.64\,\d), whilst efficient prewhitening of the 2010 data requires three frequencies (1.65, 0.82, and 0.37\,\d), which are the three strongest, non-orbital, frequencies reported by Mahy et al.\ (\cite{Mahy}). The prewhitened periodograms are shown in Fig.\,\ref{fig:TF4471}. We thus conclude that the variations in the emission humps of the secondary's \ion{He}{i} $\lambda$\,4471 line can be explained by the two dominant frequencies of the {\it CoRoT} photometry: 0.82\,\d\ and its first harmonic at 1.65\,\d, with some possible contribution of the 0.37\,\d\ frequency.

The power spectra over the absorption core are affected by the residuals from the subtraction of the primary spectrum (see Fig.\,\ref{fig:TF4471}). For the 2010 data, the corresponding mean power spectrum features essentially the same set of frequencies as for the emission humps, although with larger residuals after prewhitening. For the 2009 campaign, the power spectrum is more complex, showing considerable power at low frequencies. This is especially the case at wavelengths between 4466 and 4471\,\AA\ because the primary line remains in this wavelength domain over the entire duration of the 2009 campaign (see Fig.\,\ref{fig:dynsec}). At longer wavelengths where the moving discrete depression is seen in the dynamic spectra, the dominant frequency is near 0.4\,\d. 

\begin{figure}[ht!]
		\resizebox{8cm}{!}{\includegraphics{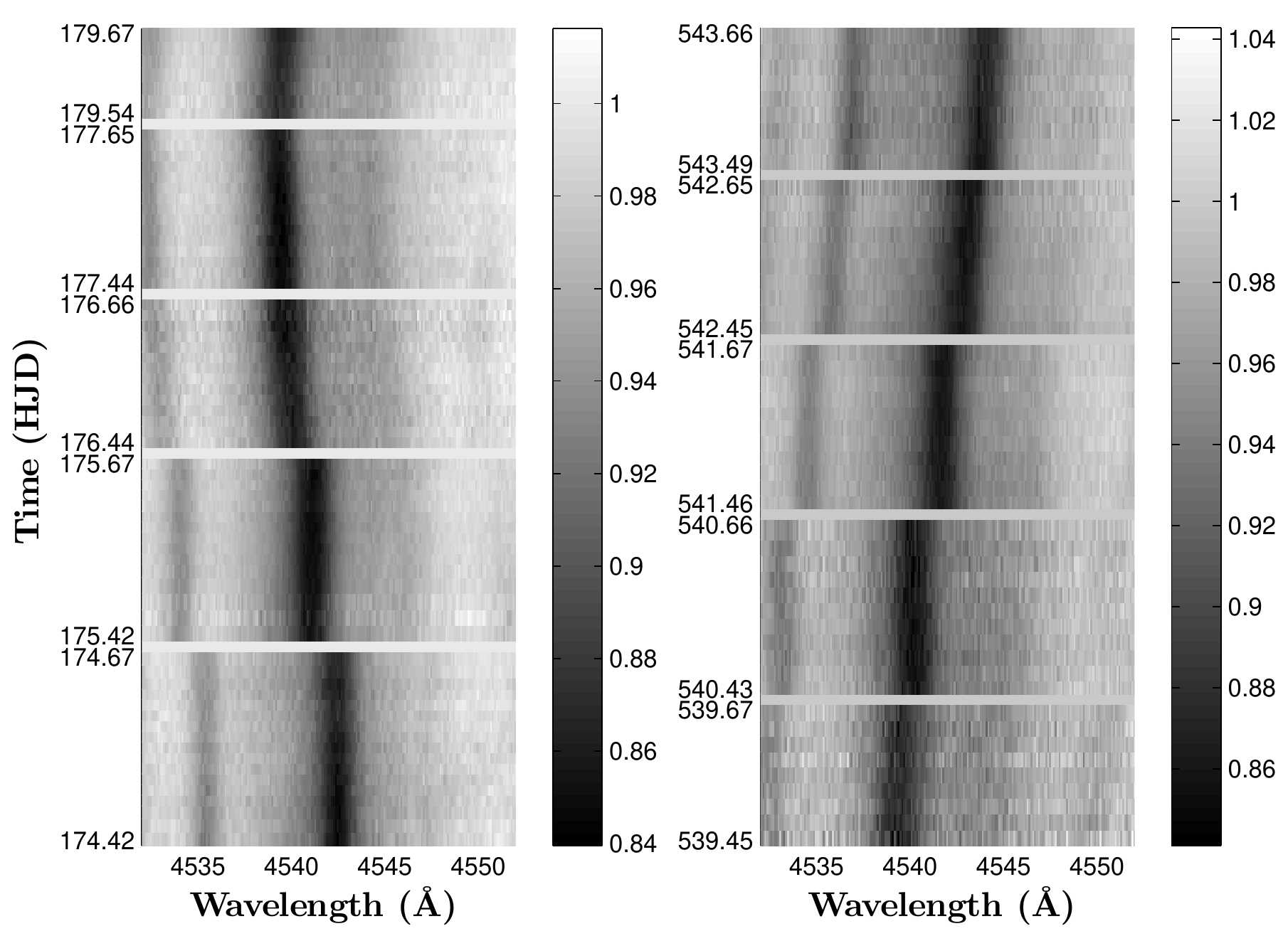}}
		\resizebox{8cm}{!}{\includegraphics{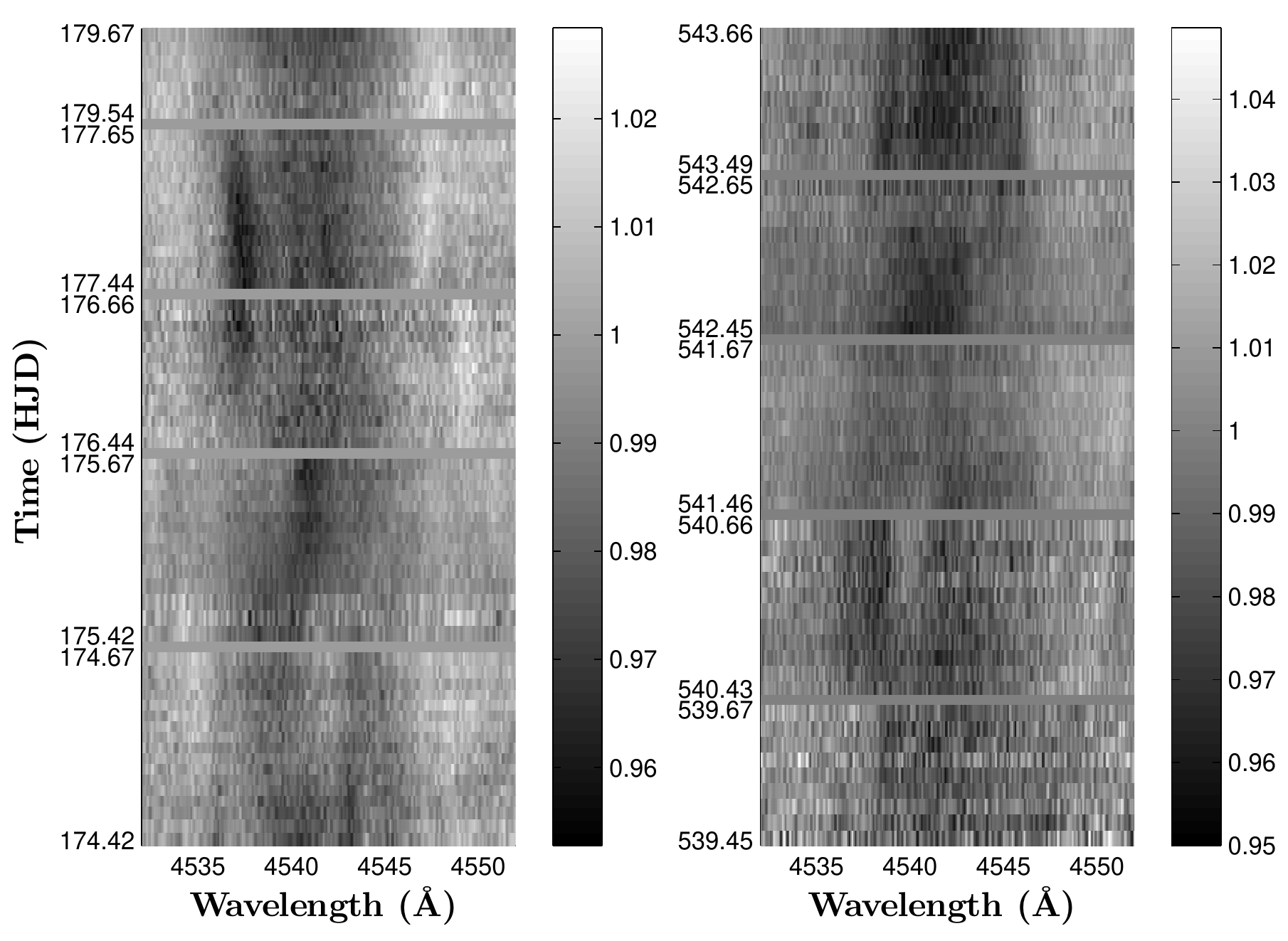}}

		\caption{Same as Fig.\,\ref{fig:dynsec}, but for the \ion{He}{ii} $\lambda$\,4542 line.}
		\label{fig:dynHeII}
\end{figure}

As pointed out above, the observed wavelengths of the primary line deviate from those expected from the orbital solution of Linder et al.\ (\cite{Linder08}). We thus performed a Fourier analysis of the \ion{He}{i} $\lambda$\,4471 line in the primary's frame of reference (adopting the Linder et al.\ \cite{Linder08} orbital solution). We restrict our 2D Fourier analysis to the core of the primary's line between 4470 and 4473\,\AA, which is roughly equivalent to the velocity range $[-v\,\sin{i},v\,\sin{i}]$. For both observing campaigns, the power spectrum is completely explained by the 0.82 and 1.65\,\d\ frequencies and their aliases. Since these are the same frequencies as those found in the variations in the secondary's emission humps, it seems quite possible that the variations are in fact due to the secondary star.

\subsection{Other lines}
The variability of the \ion{He}{ii} $\lambda$\,4542 and \ion{N}{iii} $\lambda\lambda$ 4511 -- 4518 lines is less prominent than in the case of \ion{He}{i} $\lambda$\,4471. The raw dynamic spectra of the \ion{He}{ii} and \ion{N}{iii} lines are dominated by the primary's orbital motion and do not reveal clear redwards-moving discrete depressions (see Fig.\,\ref{fig:dynHeII}). The secondary dynamic spectra are qualitatively similar to those of \ion{He}{i} $\lambda$\,4471, but the variability is more subtle. 

\begin{figure*}[ht!]
\begin{minipage}{6.6cm}
		\resizebox{\hsize}{!}{\includegraphics{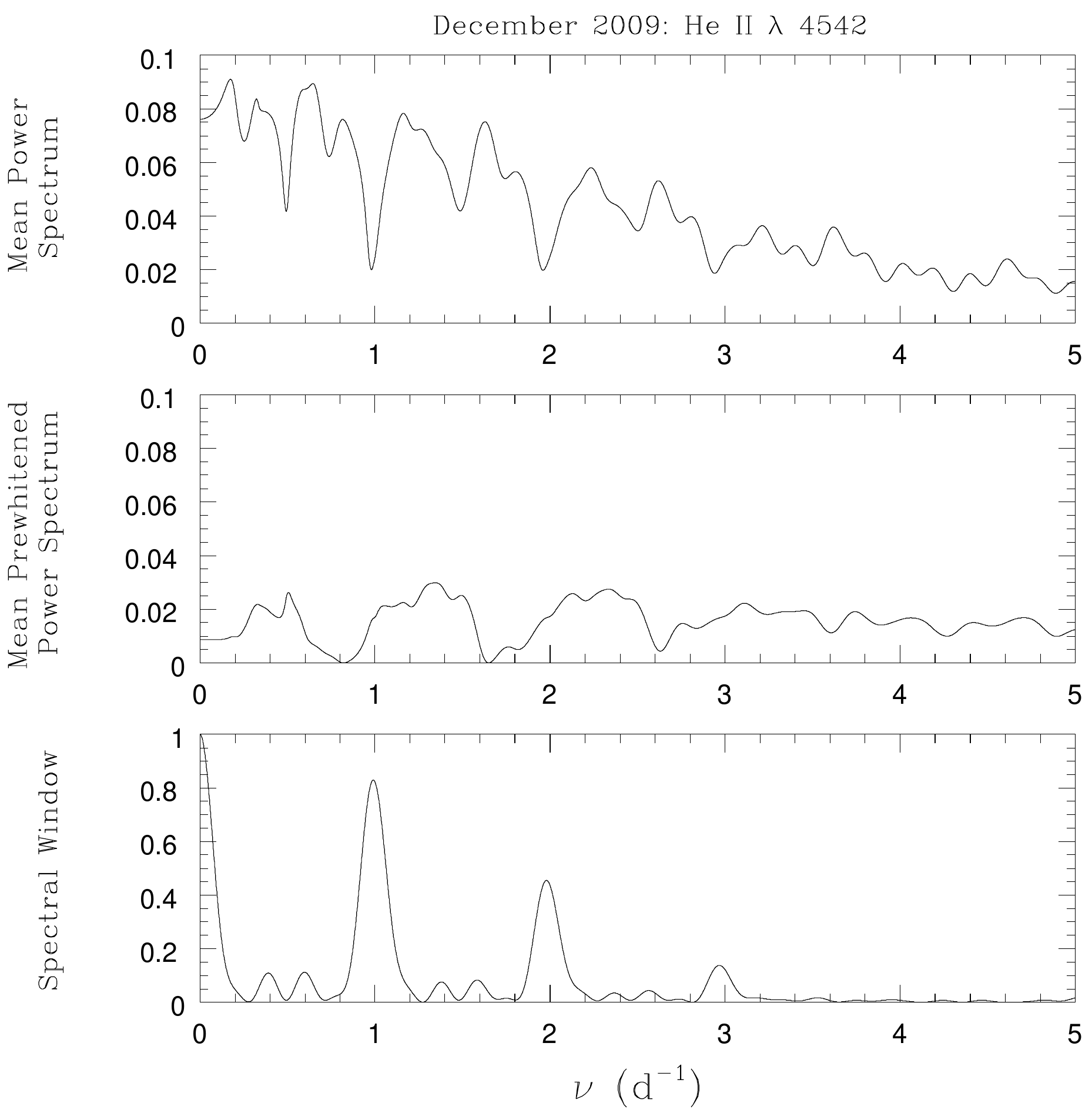}}
\end{minipage}
\begin{minipage}{6.6cm}
		\resizebox{\hsize}{!}{\includegraphics{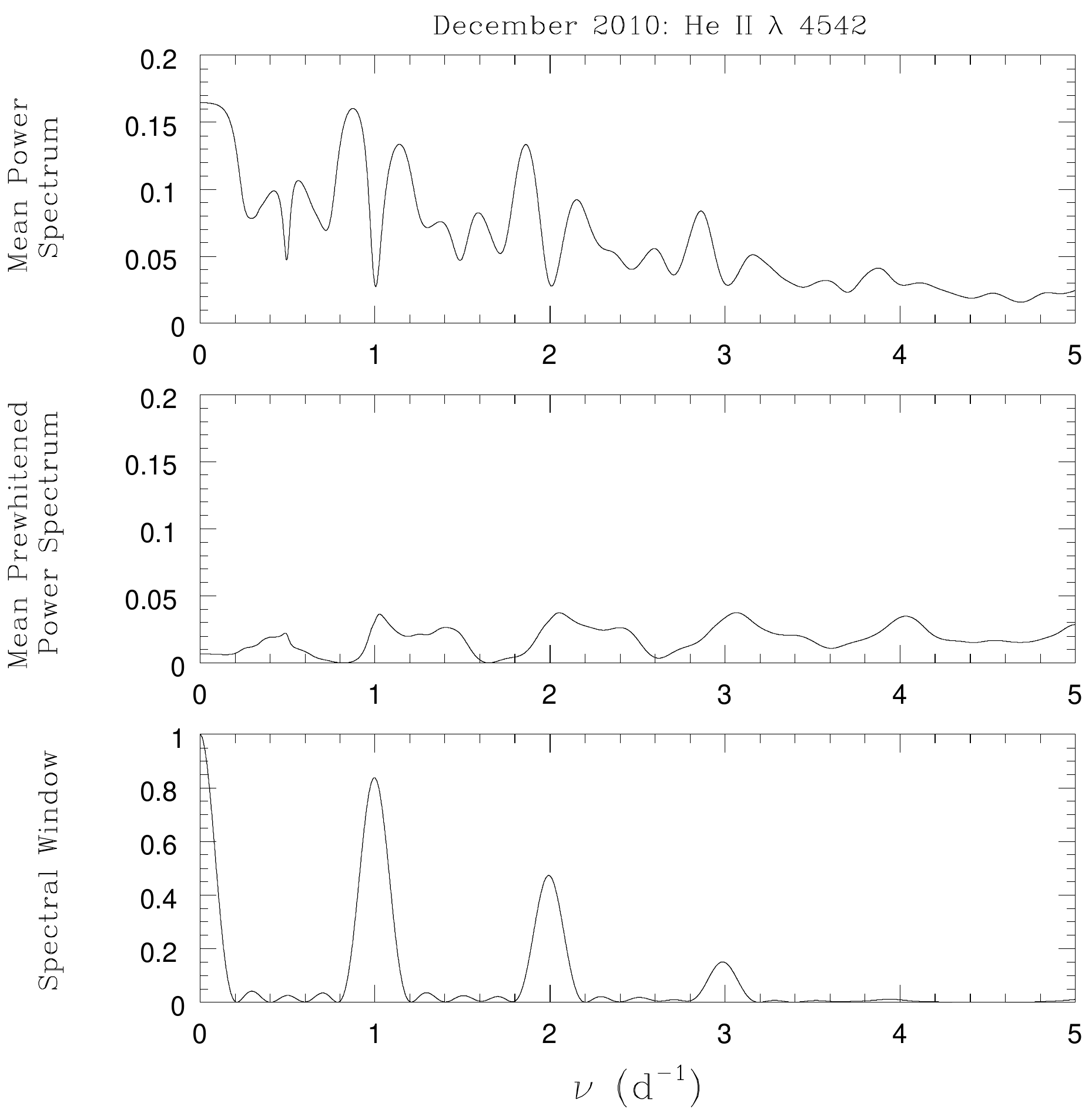}}
\end{minipage}
\begin{minipage}{5cm}
		\caption{Same as Fig.\,\ref{fig:TF4471} but for the \ion{He}{ii} $\lambda$\,4542 secondary's line. The prewhitening in the middle panels was done with the 0.82 and 1.65\,\d\ frequencies.}
		\label{fig:TF4542}
\end{minipage}
\end{figure*}

The power spectra averaged over the \ion{He}{ii} $\lambda$\,4542 line in the frame of reference of the secondary star can be described by the combination of the 0.82 and 1.65\,\d\ frequencies and their aliases (see Fig.\,\ref{fig:TF4542}). For the 2010 data, a slightly cleaner prewhitening is achieved with 0.37 instead of 1.65\,\d, and for the 2009 time series, a more efficient result is achieved including another harmonic frequency (2.47\,\d). Shifting the spectra into the frame of reference of the primary star essentially yields the same results.  

Given the nitrogen overabundance of the primary star and the possible underabundance of the secondary reported by Linder et al.\ (\cite{Linder08}), we expect a priori that the \ion{N}{iii} lines should mainly be formed in the atmosphere of the primary star\footnote{We note however that Fig.\,\ref{fig:disent} reveals some shallow \ion{N}{iii} absorptions also in the secondary spectrum.}. We have analysed the two strongest \ion{N}{iii} lines ($\lambda\lambda$\,4511 \& 4515) in the frame of reference of the primary. The resulting power spectra are again explained by the 0.82 and 1.65\,\d\ frequencies and (in the 2009 data) some contribution at 0.37\,\d. 

	\begin{table}
		\caption{Correspondence between the best frequencies (in \d) found in this paper and those found by Mahy et al. (\cite{Mahy}) in the {\it CoRoT} light curve.}
		\label{tab:corot}
		\centering
			\begin{tabular}{l c c c c}
			 \hline\hline
				Our Frequencies & \multicolumn{4}{c}{CoRoT Frequencies}\\
				\hline
				$0.37$ & 0.368 & 0.399 & (0.650) & (1.646)\\
				$0.82$ & 0.823 & 0.799 \\		
				$1.65$ & 1.646 & (0.650) & (0.368) & (0.399)\\
				\hline
			\end{tabular}
			\tablefoot{Photometric frequencies between brackets correspond to aliases due the 1-day aliasing that affects our time series.}
	\end{table}

\section{Discussion \label{sect:scen}}
The analysis of our spectroscopic time series of Plaskett's star in the previous section revealed three main frequencies 0.82, 1.65, and 0.37\,\d, although the properties of the latter two are interdependent because of the aliasing problem. Before we discuss the possible interpretation of these results, let us first recall the ingredients of Plaskett's star as (we think) we know this system. The primary star is a rather slow rotator presenting chemical enrichment indicating that it is most probably in a post-RLOF status. The secondary is a fast rotator probably surrounded by a flattened wind. The orbit is circular and the stars currently do not fill their Roche lobes. A magnetic field, probably associated with the secondary, was detected. This magnetic field could be at the origin of the flattened wind structure.

With these ingredients in mind, three scenarios could possibly explain the observed line profile variations: pulsations, tidal interactions, and an oblique magnetic rotator.

\subsection{Pulsations}
Mahy et al.\ (\cite{Mahy}) tentatively assigned the 0.82\,\d\ frequency and its six harmonics to a low-order ($2 \leq l \leq 4$) non-radial pulsation (NRP). Our Fourier analyses provide us with the amplitude ($A$) and the phase constant ($\varphi_0$) of the variation as a function of wavelength. These quantities can help us test the NRP scenario. Indeed, for NRPs with moderate degrees $l$ the phase constant usually changes monotonically across the line profile and the difference between the phase constant in the red wing and in the blue wing is directly related to the values of $l$ and $m$ of the pulsation (e.g. Schrijvers \& Telting \cite{Schrijvers}, Zima et al. \cite{Zima}).

\begin{figure*}[ht!]
\begin{minipage}{8cm}
\resizebox{\hsize}{!}{\includegraphics{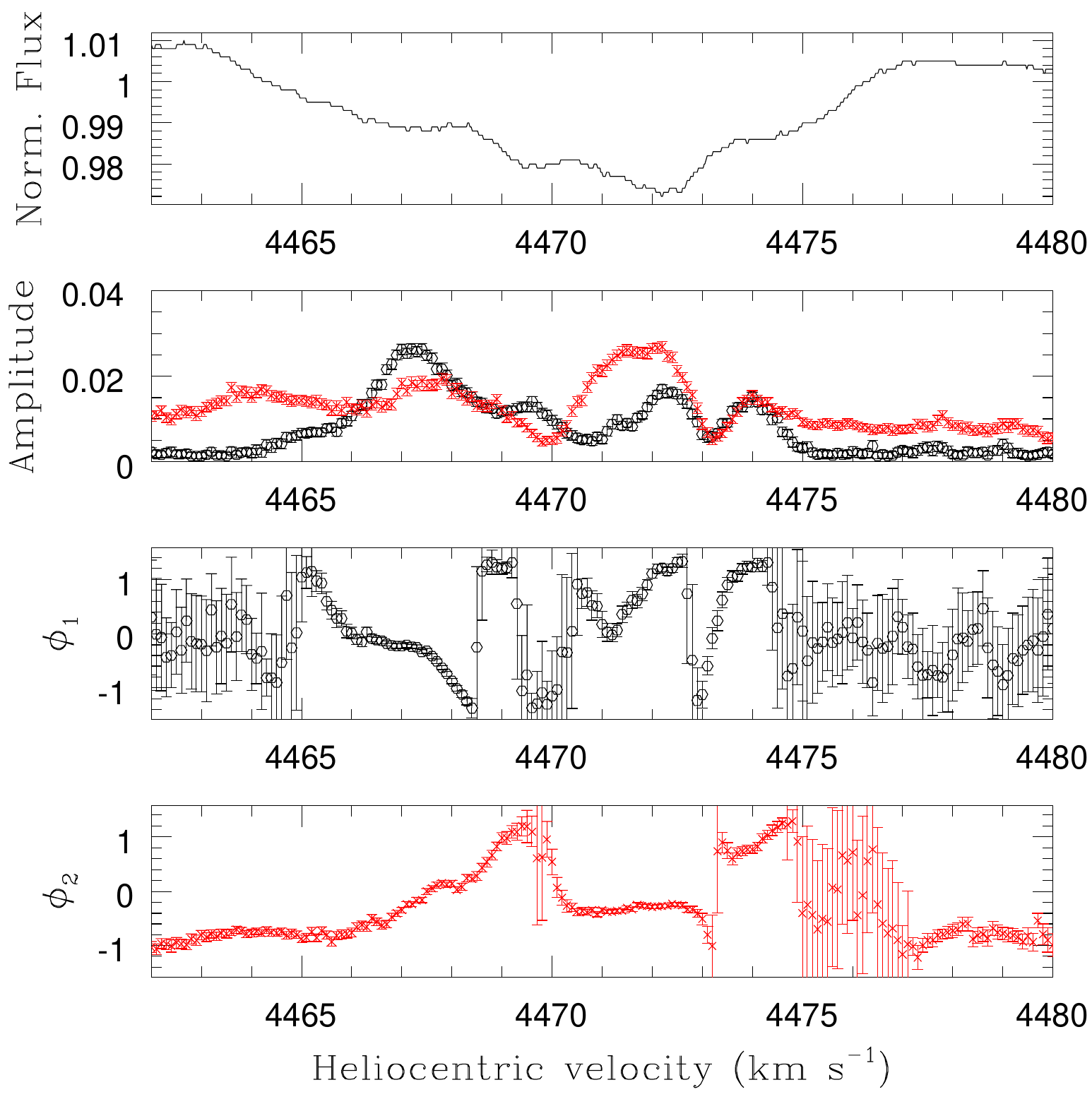}}
\end{minipage}
\begin{minipage}{8cm}
\resizebox{\hsize}{!}{\includegraphics{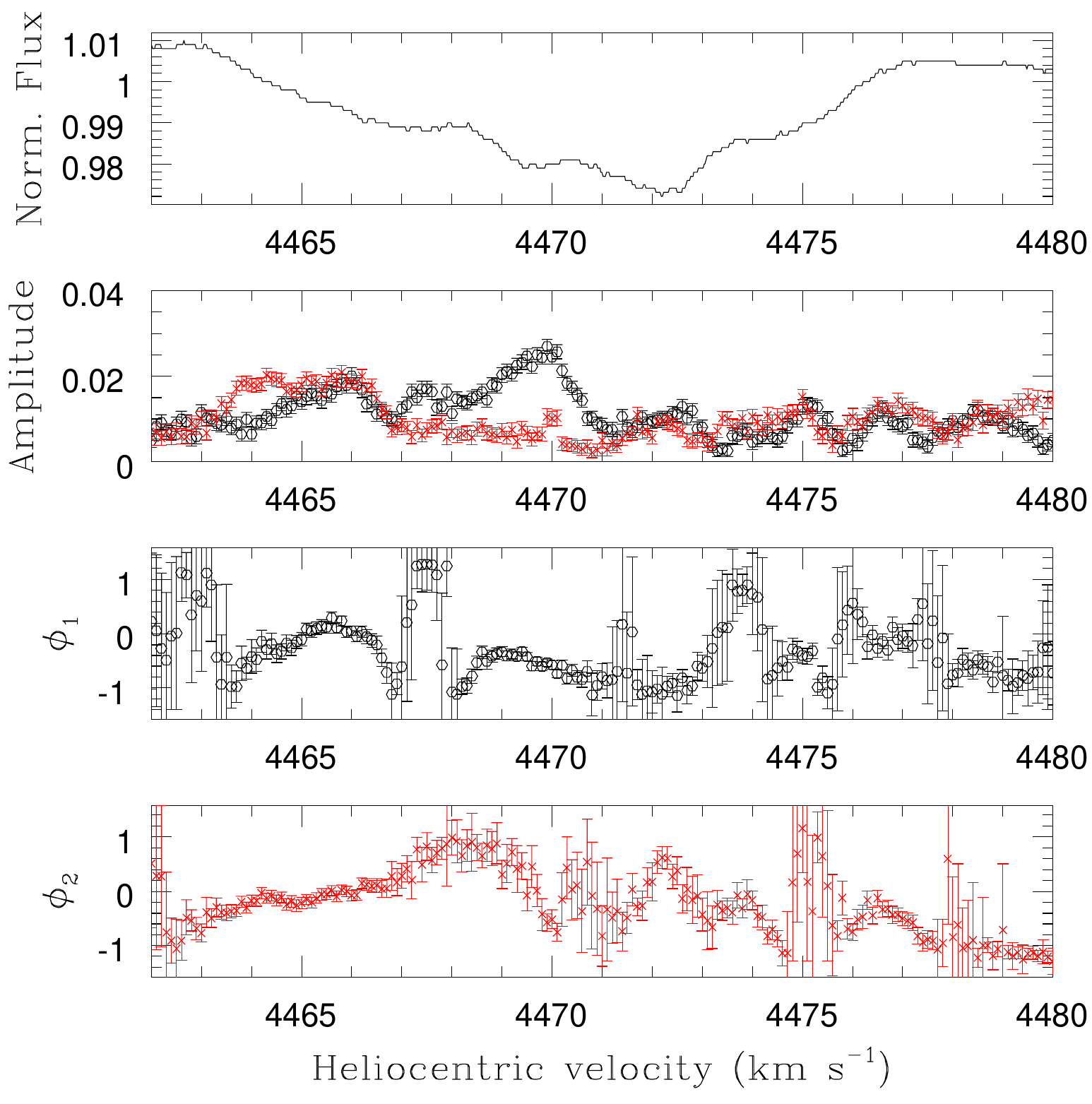}}
\end{minipage}
\caption{Amplitude and phase variations for the \ion{He}{i} $\lambda$\,4471 line in the frame of reference of the secondary during the 2009 (left) and 2010 (right) campaigns. Two frequencies have been considered simultaneously: $\nu_1 = 0.82$ and $\nu_2 = 1.65$\,\d. The top panels yield the mean secondary profile, whilst the amplitudes associated with the two frequencies are shown in the second panels. The two lower panels indicate the phase constants of these frequencies.}
\label{fig:ampli}
\begin{minipage}{8cm}
\resizebox{\hsize}{!}{\includegraphics{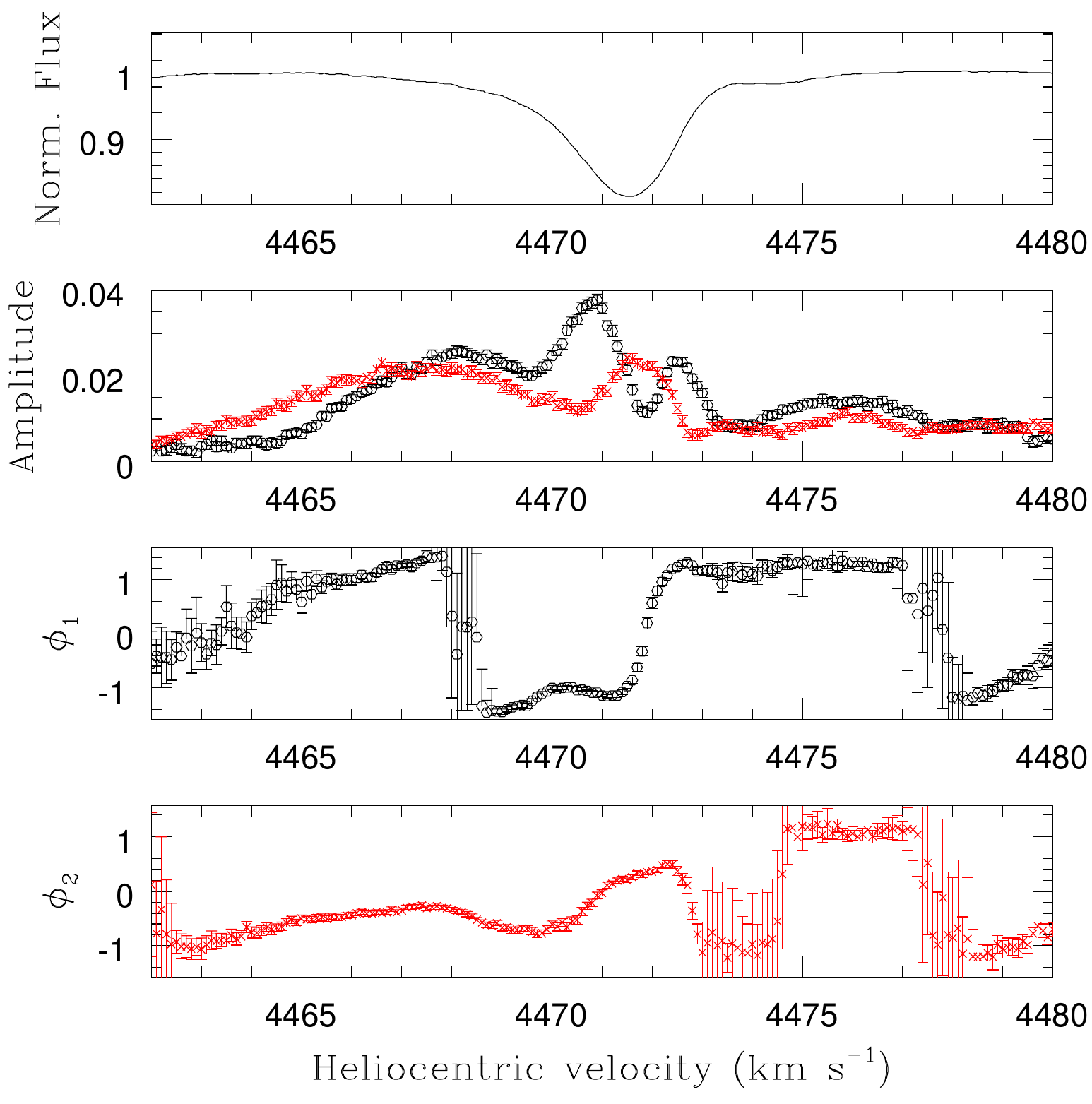}}
\end{minipage}
\begin{minipage}{8cm}
\resizebox{\hsize}{!}{\includegraphics{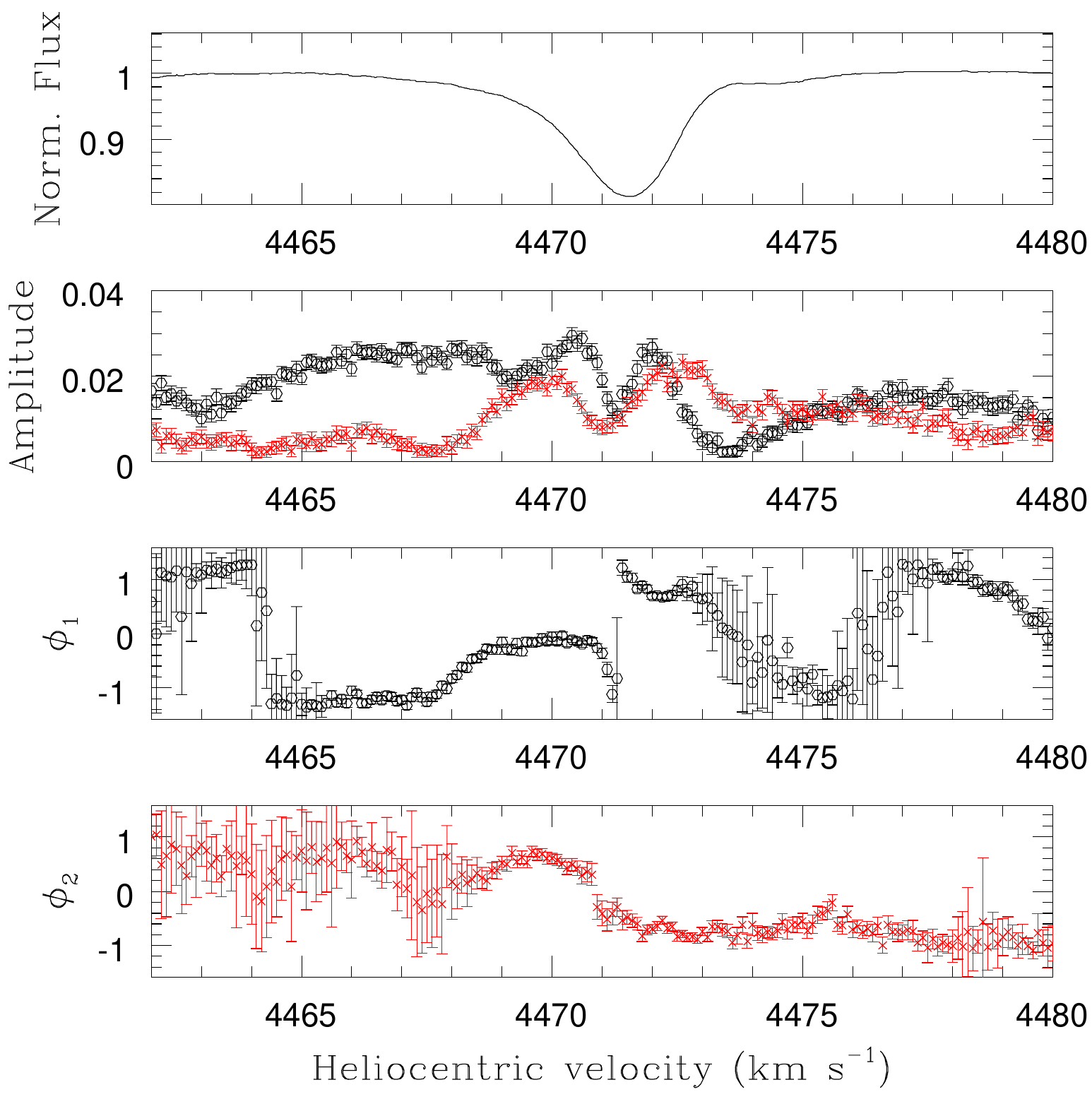}}
\end{minipage}
\caption{Same as Fig.\,\ref{fig:ampli}, but in the frame of reference of the primary.}
\label{fig:ampli2}
\end{figure*}

The amplitude and phase distributions of the 0.82 and 1.64\,\d\ frequencies of the \ion{He}{i} $\lambda$\,4471 secondary line are shown in Fig.\,\ref{fig:ampli}. The uncertainties were evaluated via Monte Carlo simulations (see Rauw et al.\ \cite{Rauw}). Whilst these distributions are possibly affected by the residuals of the primary line, they are quite erratic and do not ressemble what one would expect for typical NRPs.

If we consider the amplitudes and phase constants in the frame of reference of the primary, the variations are less erratic and, at least in the case of the 0.82\,\d\ frequency, are more reminiscent of genuine NRPs. This would imply that the primary is responsible for the 0.82 and 1.65\,\d\ frequencies. In the frame of reference of the primary, significant amplitude of variability is found far away from the core of the primary line, out to $-500$ (-6.6\,$v\sin{i}$) and $+400$\,km\,s$^{-1}$ (5.3\,$v\,\sin{i}$) away from the line centre. Similar results are obtained for the \ion{He}{ii} $\lambda$\,4542 line. Variability over such an extended wavelength range around the line core is however rather unexpected for conventional NRPs in a slow rotator, therefore suggesting that the observed variability is most likely associated with the secondary star. The apparent variability of the primary could then be induced by travelling features of the much broader secondary line when they cross the position of the primary line.

\subsection{Tidal interactions}
The analysis of the line profile variations in the slightly eccentric binary system Spica (Palate et al. \cite{PalateSpica} and references therein) has shown that tidal interactions could explain Spica's line profile variations. Although Plaskett's star is not an eccentric system, the non-synchronicity of the secondary rotation could lead to tidally induced variations such as those reported by Moreno et al. (\cite{Morenob}). Therefore, we can wonder whether tidal interactions could be responsible for the variations reported hereabove.

We have thus tried to model Plaskett's star with the TIDES code (see Moreno et al. \cite{Morenoa,Morenob,Morenoc}) combined to CoMBiSpeC\footnote{TIDES: tidal interaction with dissipation of energy through shear. CoMBiSpeC: code of massive binary spectral computation.} (see Palate \& Rauw \cite{Palate} and Palate et al. \cite{PRKM}). The TIDES code computes the time-dependent shape of the stellar surface for eccentric and/or asynchronous systems accounting for centrifugal and Coriolis forces, gas pressure, viscous effects and gravitational interactions. From that, CoMBiSpeC computes the synthetic spectra of the stars at several orbital phases.

We stress that our goal here is not to fine tune the parameters that reproduce perfectly the spectra of the components of Plaskett's star and their observed variations. Here, we rather wish to test whether or not tidal interactions can produce detectable variations in the spectra of this system. Indeed, fitting the spectra rigorously is a challenging task because of a large number of free/unknown parameters and the non-solar composition of the stars (see Linder et al.\ \cite{Linder08}); CoMBiSpeC currently only works with atmosphere models that have a solar composition. In the present paper, we have only adjusted the temperatures to reproduce the observed strength of the \he{i}{4471} and \he{ii}{4542} lines. The parameters used are listed in Table \ref{tab:param}. 

Mahy et al.\ (\cite{Mahy}) suggested that the frequencies near 0.4\,\d\ could correspond to the rotational frequency of the secondary. The projected equatorial rotational velocity of the secondary has been evaluated to $\sim300$\,\kms\ by Linder et al. \cite{Linder08}. These authors also derived a $\logg$ equal to $3.5 \pm 0.1$. Adopting an inclination of 67\degr\ as proposed by Mahy et al. (\cite{Mahy}), and considering that the rotation frequency is 0.4 \d, the projected rotational velocity of 300 \kms\ yields a radius $R = \frac{v_{\rm rot}}{2\,\pi\,\nu_{\rm rot}}$ of 16.1 \rsun\ for the secondary. The $\logg$ and minimum mass of the secondary ($m_S\,\sin^3{i} = 47.3$\,\msun, Linder et al.\ \cite{Linder08}) along with the adopted inclination of 67\degr\ (Mahy et al. \cite{Mahy}) rather suggest a radius\footnote{The latter was determined from the relation $g = \frac{GM}{R^2}\,(1 - \Gamma)$ where $\Gamma$ accounts for the radiation pressure of the star.} of $20.7_{-2.2}^{+1.3}$ \rsun. In this latter case, the projected rotational velocity would be equal to $\sim 385$\,\kms. The discrepancy between these rotational velocities can be explained by several factors: the uncertainties on the inclination (Mahy et al. \cite{Mahy} indicated a range from 30 to 80\degr), the error on the rotational velocity determination, on the $\logg$, and on the frequency.

\begin{table}[ht]
	\begin{center}
	 \caption{Parameters used for simulation with the TIDES + CoMBiSpeC model.\label{tab:param}}
		\begin{tabular}{l c c}
			\hline\hline
			Parameters & Primary & Secondary \\
			\hline\hline
			Period (day) & \multicolumn{2}{c}{$14.396$} \\
			Eccentricity & \multicolumn{2}{c}{$0$} \\
			Inclination ($\degr$) & \multicolumn{2}{c}{$67$} \\
			Mass ($M_{\sun}$)& $58.2$ & $60.6$ \\
			Polar temperature (K) & $33000$ & $32000$ \\
			Polar radius (\rsun) & $20.0$ & $20.7$ \\
			Polar log(g) (cgs) & $\simeq 3.5$ & $\simeq 3.5$ \\
			${\mathrm v}_{rot}$ (\kms) & $70.3$ & $296$ \\
			${\mathrm v}\sin i$ (\kms) & $64.7$ & $273$\\
			$\beta^\text{ (a)}$ & $1.0$ & $4.12$\\
			Viscosity, $\nu$ (\rsun$^2$\d) & $0.028$ & $0.035$ \\
			Layer depth ($\Delta R/R$) & $0.07$ & $0.07$\\
			Polytropic index & $1.5$ & $1.5$ \\
			Number of azimuthal partitions & $500$ & $500$\\
			Number of latitudinal partitions& $20$ & $20$\\
	  	\hline
		\end{tabular}
		\end{center}
 	\tablefoot{$^\text{(a)}$The $\beta$ parameter measures the asynchronicity of the star at periastron and is defined by $\beta = 0.02\frac{Pv_{rot}}{R}\times \frac{(1-e)^{3/2}}{(1+e)^{1/2}}$, where $v_{rot}$ is the equatorial rotation velocity, $R$ is the equilibrium radius, and $e$ is the eccentricity.}
	\end{table}

\begin{figure}[ht!]
		\resizebox{\hsize}{!}{\includegraphics{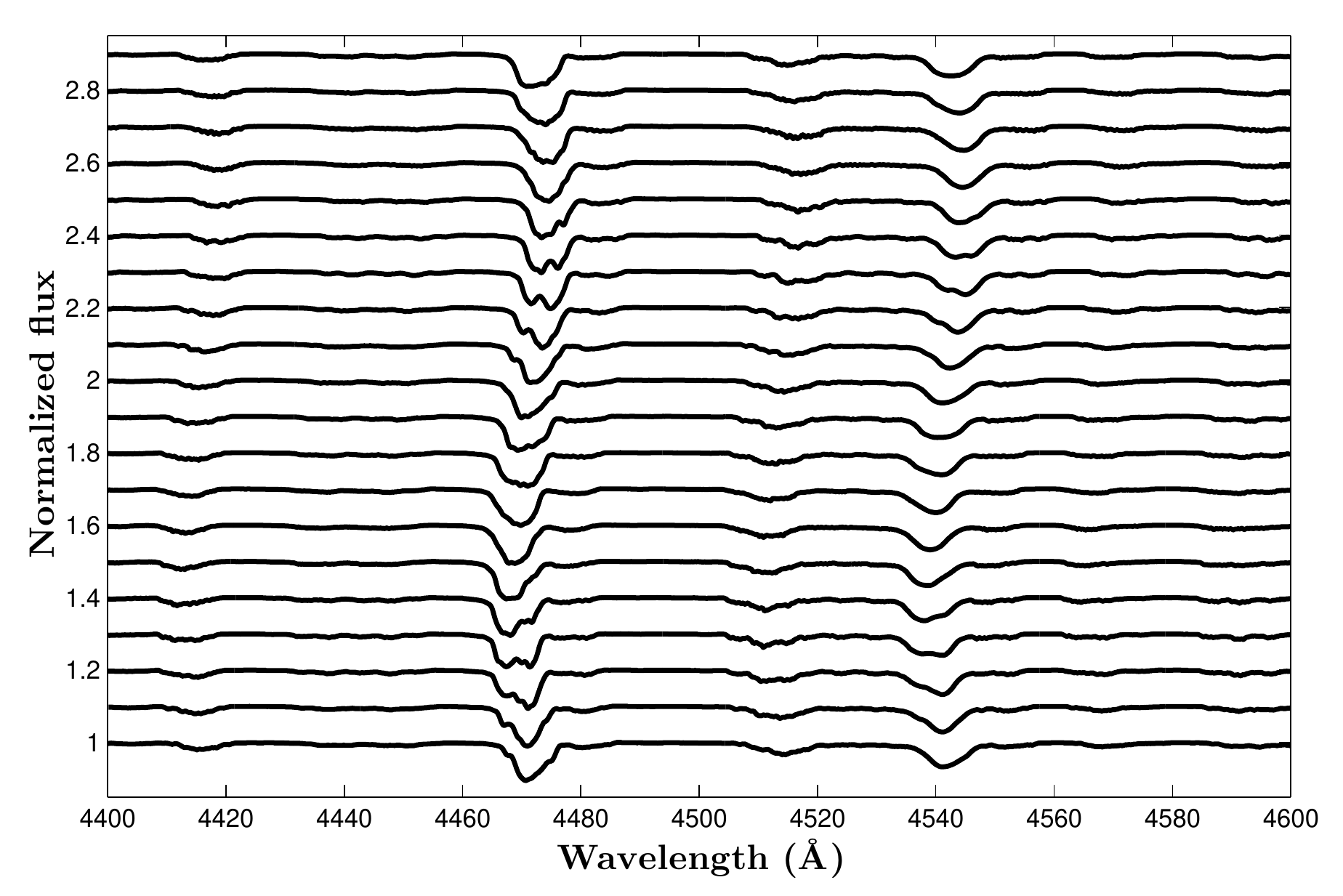}}
		\caption{Line profile variations in the synthetic spectra of the secondary during the orbital cycle. The spectra (separated by 0.05 in phase) have been shifted vertically by 0.1 continuum unit for clarity.}
				\label{fig:modS}
\end{figure}

Our simulations indicate that tidal interactions can produce visible variations in the lines of the secondary. Because the primary is thought to be in synchronous rotation in a circular orbit, it is expected that there is no variability of the primary due to tidal interactions. The strength of the simulated variations is comparable to the observed ones. Reducing the secondary radius to 18\,\rsun\ slightly lowers the amplitude of the tidally induced variations, although they remain well visible in the synthetic spectra. Fig.\ \ref{fig:modS} displays the simulated line profile variations in the secondary during the orbital cycle. Whilst our code cannot simulate the impact of the tidal interactions on a co-rotating confined wind, it seems very plausible that the tidal interactions also affect the latter and thus extend into the emission humps. This result brings up the interesting possibility that breaking tidal waves near the secondary's equator could produce an enhanced mass-loss in this region (see Osaki \cite{Osaki}). Combined with wind confinement due to the magnetic field (Grunhut et al.\ \cite{Grunhut,Grunhut2}), this effect could produce the equatorial torrus of circumstellar material that was advocated by Wiggs \& Gies (\cite{Wiggs}) and Linder et al.\ (\cite{Linder06, Linder08}). The much smaller radius (10.7\,\rsun) proposed by Grunhut et al. (\cite{Grunhut}) not only fails to reproduce the $\log{g}$ value inferred from the model atmosphere code fits (see below), but would also considerably reduce the amplitude of surface oscillations due to tidal interactions.

We have performed a Fourier analysis of the individual simulated spectra of the secondary stars. The power spectrum displays a peak at the secondary's rotation frequency imposed in the model, but does not show peaks at the harmonic frequencies. Tidal interactions can thus induce profile variations, provided the stellar radii are sufficiently large. However, the current model does not explain the full frequency content of HD\,47129's spectroscopic and photometric variations.

\subsection{Magnetically confined winds}
Grunhut et al.\ (\cite{Grunhut}) reported on the discovery of a magnetic field in Plaskett's star with a minimum surface dipolar strength of 2.8\,kG at the magnetic poles. Such a strong magnetic field could confine the secondary's stellar wind, thereby producing the emission humps seen around the \ion{He}{i} $\lambda$\,4471 line as well as the double-peaked H$\alpha$ emission. Using additional spectropolarimetric observations, Grunhut et al.\ (\cite{Grunhut2}) inferred a tilt angle of the magnetic axis of $(80 \pm 5)^{\circ}$ with respect to the rotation axis. This would result in a magnetic configuration very similar to HD\,57682 (Grunhut et al.\ \cite{Grunhut3}). One would then expect a double-wave modulation of the strength of the emission features. Based on a periodicity search in the variations in the H$\alpha$ equivalent width and the longitudinal magnetic field, Grunhut et al.\ (\cite{Grunhut2}) inferred a rotational period of 1.215\,d for the secondary, corresponding to the 0.82\,\d\ frequency.

This scenario could thus explain our detection of the 0.82 and 1.65\,\d\ frequencies quite naturally in the emission humps. However, adopting 0.82\,\d\ as the secondary's rotational frequency would lead to a radius estimate from the projected rotational velocity that is twice smaller than what we have obtained above, thus enhancing the discrepancy between the radius estimated from $v\,\sin{i}$ and the one estimated from $\logg$. 

Concerning the latter discrepancy, we note that Wiggs \& Gies (\cite{Wiggs}) previously reported a 2.78\,d period in the equivalent width of the H$\alpha$ emission wings\footnote{Wiggs \& Gies \cite{Wiggs} suggested an origin in the winds of the stars, corresponding to a frequency of 0.36\,\d\ in good agreement with the 0.37\,\d\ rotation frequency suggested by Mahy et al.\ (\cite{Mahy}). A corollary would be that the variations in the longitudinal magnetic field presented by Grunhut et al.\ (\cite{Grunhut2}) would probably have to be interpreted as due to a more complex magnetic field than a simple dipole.}.

\subsection{Summary and conclusion}

Our analysis of the \ion{He}{i} $\lambda$\,4471, \ion{N}{iii} $\lambda\lambda$\,4510-4518, and \ion{He}{ii} $\lambda$\,4542 lines of Plaskett's star revealed variations at several frequencies: 0.37, 0.82, and 1.65\d. The strongest variations are found for the \ion{He}{i} line.

Considering the current knowledge of the system, it is difficult to assign the spectroscopic variability to either the primary or the secondary star. Indeed, our analysis yielded conflicting results in this respect: whilst the velocity range over which variations are detected clearly favours the secondary star, the pattern of the amplitude and phase of the variations makes more sense if they originate in the primary star. As a result, none of the current scenarios (pulsations, tidal interactions, magnetically confined winds) accounts for all the aspects of the observed variations.

\begin{acknowledgements}
We would like to thank Pr.\ G.\ Koenigsberger who provided us with the TIDES code and helped us to use it. We are grateful to the referee, Dr.\ Jason Grunhut who helped us improve this paper. We acknowledge support through the XMM/INTEGRAL PRODEX contract (Belspo), from the Fonds de Recherche Scientifique (FRS/FNRS), as well as by the Communaut\'e Fran\c caise de Belgique - Action de recherche concert\'ee - Acad\'emie Wallonie - Europe.
\end{acknowledgements}


\end{document}